\documentclass{article}
\pdfoutput=1
\usepackage{spconf,amsmath,graphicx}
\usepackage[ansinew]{inputenc}
\usepackage[english]{babel}
\usepackage[T1]{fontenc}
\usepackage{psfrag}
\usepackage{hyperref}
\usepackage{multirow}
\usepackage{units}
\usepackage{placeins}
\usepackage{microtype}
\usepackage{color}
\usepackage{cite}

\makeatletter
\def\@name{ \emph{Ahmed Hussen Abdelaziz$^{1,5}$\sthanks{This work was carried out at ICSI, supported by Federal Ministry of Education and Research (BMBF) through the FITweltweit program, administered by German Academic Exchange Service (DAAD).}, Shuo-Yiin Chang$^{1\dag}$, Nelson Morgan$^{1\dag}$, Erik Edwards$^{1,4}$\sthanks{This work was partially supported by National Science Foundation grant IIS:1320260 and grant IIS:1320366.},}  \\ 
\emph{Dorothea Kolossa$^2$, Dan Ellis$^{1,3}$, David A. Moses$^4$, Edward F. Chang$^4$\sthanks{This work was partially supported by Institutes of Health grant R01-DC012379, New York Stem Cell Foundation, and McKnight Foundation.}}\vspace{1mm}}
\makeatother

\title{On Neural Phone Recognition of Mixed-source ECOG signals}
\vspace{-18pt}
\address{ 
$^1$International Computer Science Institute, Berkeley, CA, USA \\
$^2$ Cognitive Signal Processing Group, Ruhr-Universit{\"a}t Bochum, Germany \\
$^3$ Electrical Engineering Department, Columbia University NY, NY, USA \\
$^4$ Department of Neurological Surgery, UC San Francisco, San Francisco, CA, USA \\ 
$^5$ Apple Inc., Cupertino, CA, USA \vspace{0.5mm} \\ 
  \{ahmedha, shuoyiin, morgan\}@icsi.berkeley.edu, erik.edwards4@gmail.com, \\
dorothea.kolossa@rub.de, dpwe@ee.columbia.edu, \{david.moses, edward.chang\}@ucsf.edu}

\begin{document}
%\ninept

\maketitle
\vspace{-5pt}
\begin{abstract}
The emerging field of neural speech recognition (NSR) using electrocorticography has recently attracted remarkable research interest for studying how human brains recognize speech in quiet and noisy surroundings. In this study, we demonstrate the utility of NSR systems to objectively prove the ability of human beings to attend to a single speech source while suppressing the interfering signals in a simulated cocktail party scenario. The experimental results show that the relative degradation of the NSR system performance when tested in a mixed-source scenario is significantly lower than that of automatic speech recognition (ASR). In this paper, we have significantly enhanced the performance of our recently published framework by using manual alignments for initialization instead of the flat start technique. We have also improved the NSR system performance by accounting for the possible transcription mismatch between the acoustic and neural signals.
\end{abstract}

\vspace{-5pt}
\section{Introduction}
\label{sec:intro}

In the realm of automatic speech recognition (ASR), the ultimate goal is to simulate or even to exceed the ability of human brains in transforming the acoustic stimuli into a set of fundamental speech units (phones, syllables, words) whose composite can give a meaningful sentence. Modern ASR systems can perform this task quite well in quiet surroundings. However, in noisy environments, the performance of ASR systems is poor compared to human beings, who can easily attend to a single sound source and suppress the other interfering sounds. 

This phenomenon has been objectively investigated in \cite{mesgarani2012} by reconstructing the spectrotemporal representation of speech from electrocorticographic (ECoG) signals\footnote{Signals recorded directly from the brain surface using an electrode array.}\cite{blakely2008localization,chang2010categorical,chang2011cortical,pei2011decoding,mugler2014direct,herff2015brain,heger2015continuous} that were recorded while human subjects were listening to single- and mixed-source speech signals. In \cite{Chang2015}, model-based techniques have been used to objectively demonstrate the same phenomenon. The idea was to train a so-called neural speech recognition (NSR) system \cite{moses2016neural} using single-source ECoG signals. The NSR system was then tested using single- and mixed-speaker ECoG signals. The results were finally compared to ASR performance in the same testing scenarios.

In this paper we also use an NSR system to \emph{objectively} demonstrate the fact that humans are less affected by interfering sounds in cocktail party scenarios than ASR systems. 

We have modified our previous framework in \cite{Chang2015} so that significant improvements have been achieved. One of the main modifications that has improved both the ASR and NSR results is the change of the initialization technique of the ASR models. In this study, we have initialized the ASR models with a set of phone-based manually transcribed utterances instead of the flat start technique used in \cite{Chang2015}.

The NSR results have also been improved by accounting for the fact that the transcription and the frame-state alignments of the neural signals may differ from those of the acoustic signals. We have investigated this fact by adding a silence after every phone in the training lattices of the NSR system with 0.9 silence skip probability. This modification has allowed for silences that occur only in the ECoG signals to be consumed in the newly embedded silence path and not to be involved in the training of the neural phone model. 

The remaining paper is organized as follows: In Section \ref{sec:NSR}, an overview of the framework is given. Next, in Section \ref{sec:ANA}, the alignment mismatch problem is discussed. In Section \ref{sec:Experiments and Results}, the framework is evaluated in a single- and mixed-speaker scenarios. Finally, the paper is concluded and an outlook of future work is given in Section \ref{sec:Conclusions}.

%\section{Relation }
%Training NSR systems for continuous phone recognition is a challenging task because of the small amount of the available training data. In \cite{}, an NSR system has been trained to decode ECoG signals recorded during speech production task. In \cite{}, a Gassian mixture model/hidden Markov model (GMM)/(HMM) NSR system has been trained to decoded ECoG signals recorded during speech perception. Other type of classifiers such as linear and Bayes classifiers and support vector machine (SVM) have also been used in prior works to decoded isolated phones or words using other types of brain
%imaging techniques, see for example \cite{}.  

%We have further investigated the possible differences between the ASR and the NSR transcription by using the so-called coupled HMM (CHMM) \cite{Abdelaziz2014d}. The CHMM is a multimodal fusion model that takes into account the asynchronicities between different modalities while enforcing the synchronization at sentence boundaries. The CHMM has been used to filter the NSR training set by choosing only the neural training utterances whose transcription mostly match the transcription of the acoustic utterances.

%%%%%%%%%%%%%%%%%%%%%%%%%%%%%%%%%%%%%%%%%%%%%%%%%%%%%%%%%%%%%%%%%%%%%%%%
												%																															Figures/				 %
												%					    											Framework												 %
												%																																			 %
												%%%%%%%%%%%%%%%%%%%%%%%%%%%%%%%%%%%%%%%%%%%%%%%%%%%%%%%%%%%%%%%%%%%%%%%%
\section{Framework}
\label{sec:NSR}
\begin{figure}[t!]
%\psfrag{A}[cc][cc][0.7]{\begin{tabular}{@{}c@{}}
%   TIMIT \\
%	  CRM
%\end{tabular}}%
%\psfrag{B}[cc][cc][0.7]{\begin{tabular}{@{}c@{}}
%   ECoG \\
%	 data \\
%	 acquisition 
%\end{tabular}}%
%\psfrag{C}[cc][cc][0.7]{\begin{tabular}{@{}c@{}}
%   Neural \\
%	feature \\
%	 extraction
%\end{tabular}}%
%		\psfrag{D}[cc][cc][0.7]{NSR}
%		\psfrag{G}[cc][cc][0.7]{\begin{tabular}{@{}c@{}}
%   NSR \\
%	 decoder
%\end{tabular}}%
%		\psfrag{E}[cc][cc][0.7]{\begin{tabular}{@{}c@{}}
%   Acoustic \\
%	feature \\
%	 extraction
%\end{tabular}}%
%		\psfrag{F}[cc][cc][0.7]{ASR}
%		\psfrag{H}[cc][cc][0.7]{\begin{tabular}{@{}c@{}}
%   ASR \\
%	 decoder
%\end{tabular}}%
%		\psfrag{I}[cc][cc][0.7]{\begin{tabular}{@{}c@{}}
%   Language \\
%	 model
%\end{tabular}}%
%		\psfrag{J}[cc][cc][0.7]{\begin{tabular}{@{}c@{}}
%   Training \\
%	 set
%\end{tabular}}%
%		\psfrag{K}[cc][cc][0.7]{\begin{tabular}{@{}c@{}}
%   Test set
%	\end{tabular}}%
%	\psfrag{L}[cc][cc][0.7]{\begin{tabular}{@{}c@{}}
%   Frame-state \\
%	 alignments
%	\end{tabular}}%
	\centering
		\includegraphics[width=\linewidth]{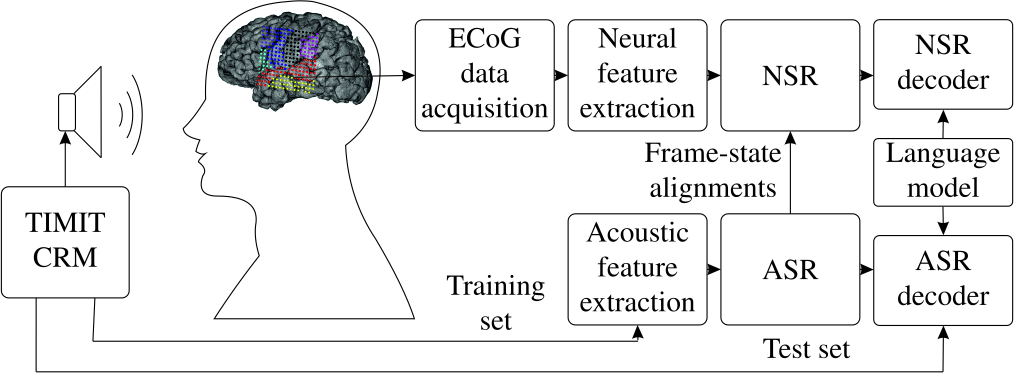}
	\caption{Framework for synchronous training/testing of neural speech recognition (NSR) and automatic speech recognition (ASR).
    \vspace{-5mm}}
	\label{fig:NSR}
\end{figure}
Fig.~\ref{fig:NSR} shows the framework used to examine the utility of the NSR systems to decode perceived speech in single- and mixed-source scenarios. As can be seen, ECoG signals are recorded from an electrode array that is superimposed directly over the brain, specifically over the auditory cortex, of a human subject. The ECoG signals were simultaneously recorded while the subject was listening to single- and mixed-source utterances from the TIMIT \cite{J1993} and CRM \cite{RS2000} corpus. Acoustic and ECoG features are then extracted from the training/testing acoustic and ECoG signals. The extracted acoustic features are used to train an ASR system, which is then used to provide the NSR system with frame-state alignments. After the ASR and NSR system are trained, they are used to recognize context-independent monophones from single- and mixed-source utterances of the test set. A biphone language model is used for phone recognition.

												%%%%%%%%%%%%%%%%%%%%%%%%%%%%%%%%%%%%%%%%%%%%%%%%%%%%%%%%%%%%%%%%%%%%%%%%
												%																																			 %
												%					    	CHMM-based Utterance Filtering												 %
												%																																			 %
												%%%%%%%%%%%%%%%%%%%%%%%%%%%%%%%%%%%%%%%%%%%%%%%%%%%%%%%%%%%%%%%%%%%%%%%%
                                        \section{Acoustic and Neural Data Alignments}
\label{sec:ANA}
The ASR system can be easily trained with any publicly available large speech corpus as done in \cite{herff2015brain,heger2015continuous}.  However, in order to conduct a fair comparison between the ASR and NSR results, we have trained the ASR system only with the acoustic utterances for which corresponding ECoG signals exist. 

\begin{figure}[t!]
%\psfrag{a}[cc][cc][0.7]{$0.2$}%
%\psfrag{b}[cc][cc][0.7]{$0.4$}%
%\psfrag{c}[cc][cc][0.7]{$0.6$}%
%\psfrag{d}[cc][cc][0.7]{$0.8$}%
%\psfrag{e}[cc][cc][0.7]{$1$}%
%\psfrag{f}[cc][cc][0.7]{$1.2$}%
%\psfrag{g}[cc][cc][0.7]{$1.4$}%
%\psfrag{h}[cc][cc][0.7]{$1.6$}%
%\psfrag{I}[cc][cc][0.7]{$1.8$}%
%
%\psfrag{j}[cc][cc][0.7]{$0$}%
%\psfrag{k}[cc][cc][0.7]{$1$}%
%\psfrag{l}[cc][cc][0.7]{$2$}%
%\psfrag{m}[cc][cc][0.7]{$3$}%
%\psfrag{n}[cc][cc][0.7]{$4$}%
%\psfrag{p}[cc][cc][0.7]{$5$}%
%\psfrag{q}[cc][cc][0.7]{$6$}%
%\psfrag{r}[cc][cc][0.7]{$7$}%
%\psfrag{s}[cc][cc][0.7]{$8$}%
%
%\psfrag{t}[cc][cc][0.7]{Neural feature index}%
%\psfrag{u}[cc][cc][0.7]{Frequency [kHz]}%
%\psfrag{v}[cc][cc][0.7]{Frequency [kHz]}%
%
%\psfrag{x}[cc][cc][0.7]{(a)}%
%\psfrag{y}[cc][cc][0.7]{(b)}%
%\psfrag{z1}[cc][cc][0.7]{(c)}%
%\psfrag{w}[cc][cc][0.7]{Time [s]}%
%
%\psfrag{z2}[cc][cc][0.7]{5}%
%\psfrag{z3}[cc][cc][0.7]{10}%
%\psfrag{z4}[cc][cc][0.7]{15}%
%\psfrag{z5}[cc][cc][0.7]{20}%
%\psfrag{z6}[cc][cc][0.7]{25}%
%\psfrag{z7}[cc][cc][0.7]{30}%
%\psfrag{z8}[cc][cc][0.7]{35}%
%\psfrag{z9}[cc][cc][0.7]{40}%
%\psfrag{z10}[cc][cc][0.7]{45}%
	\centering
		\includegraphics[width=\linewidth]{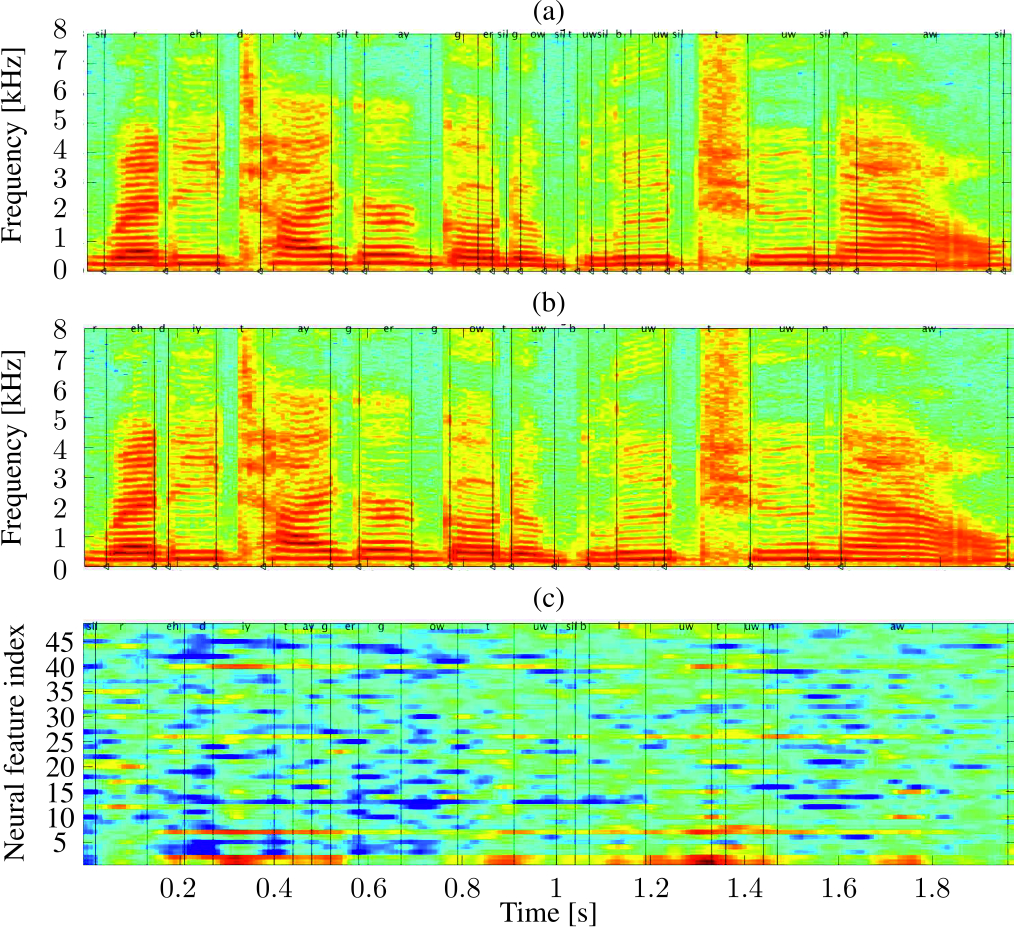}
	\caption{(a) Phone alignments obtained from an ASR system initialized with flat start. (b) Phone alignments obtained from an ASR system initialized with a set of manually transcribed data. (c) Phone alignments obtained from an NSR system.}
	\label{fig:Ali}
    \vspace{-5mm}
\end{figure}

Because of the limited amount of training data, the ASR system has to be carefully initialized in order to achieve satisfactory results. In \cite{Chang2015}, the ASR system was initialized using a flat start technique. Fig.~\ref{fig:Ali}-a shows the forced-aligned phone transcription of the CRM utterance ``Ready Tiger Go To Blue Two Now'' obtained using this ASR model. As can be seen, the ASR model falsely learns the boundaries of each phone, see, e.g., the boundaries of the phones ``r'''' and ``e'', and thus, degraded ASR results were obtained.

 In order to enhance the ASR results, we have manually transcribed a subset of the CRM corpus. The manually transcribed data has been used to initialize the ASR models. As can be seen in Fig.~\ref{fig:Ali}-b, this initialization technique has led to better phone alignments. Consequently, the ASR results have been significantly improved as will be shown in Section \ref{Results}.

As will be described in Section \ref{Models}, we use the ASR alignments to train the NSR system. Fig.~\ref{fig:Ali}-c shows the NSR forced-aligned data after the NSR training. Unlike the spectrotemporal representation of the acoustic signal, the quality of the NSR alignments can not be judged by visual inspection. However, one interesting observation can be made in Fig.~\ref{fig:Ali}-c: There are two regions in the spatiotemporal representation of the neural signal where the neural activities are more intensive than in the other regions. Those intensive activities happen when the subject attends to the call sign, here ``Tiger'', and to the color-number combination, here ``Blue Two.'' As will be described in Section \ref{CPSS}, the subject was instructed to report the color-number combination when a certain sign is called. 
This observation leads to the following question: Do the acoustic and neural transcriptions match? In this study, we investigate one aspect of this question, which is the mismatch of the silence position in the acoustic and neural transcriptions. We have added \emph{optional} silences after every phone in the neural transcription. We have also added a silence skip path in the NSR training lattices with a skip probability of 0.9. 

The idea is: if the neural signal duration of a particular phone is smaller than the corresponding acoustic signal duration, the additional silences will consume this duration mismatch and it will not wrongly be involved in training the neural phone model. As a result of this modification, additional silences appear in the NSR alignments, which do not exist in the acoustic alignments, as can be seen in Fig.~\ref{fig:Ali}. 

												%%%%%%%%%%%%%%%%%%%%%%%%%%%%%%%%%%%%%%%%%%%%%%%%%%%%%%%%%%%%%%%%%%%%%%%%%
												%%																																			 %
												%%					    	CHMM-based Utterance Filtering												 %
												%%																																			 %
												%%%%%%%%%%%%%%%%%%%%%%%%%%%%%%%%%%%%%%%%%%%%%%%%%%%%%%%%%%%%%%%%%%%%%%%%%
%\section{CHMM-based Training Set Filtering}
%\label{sec:CHMM}
%
%\section{Acoustic and Neural Data Alignments}
%\label{sec:ANA}
%\begin{figure}[t!]
%\psfrag{A}[cc][cc][0.6]{Neural model for phone ``r''}%
%\psfrag{B}[cc][cc][0.6]{Neural model for phone ``sil''}%
%\psfrag{C}[cc][cc][0.6]{Acoustic model for phone ``r''}%
%\psfrag{D}[cc][cc][0.6]{Restricted coupled HMM}%
%\psfrag{E}[cc][cc][0.6]{Audio Marginal HMM}%
%\psfrag{N}[cc][cc][0.6]{Neural marginal HMM}%
	%\centering
		%\includegraphics[width=0.7\linewidth]{Figures/CHMM.eps}
	%\caption{.}
	%\label{fig:CHMM}
%\end{figure}

%Since there are an enormous amount of publicly available acoustic data, a very-well trained ASR system could be trained. However, 

%Instead of using the acoustic alignments to directly train the deep neural network (DNN)/HMM NSR system, we have firstly used the acoustic alignments to train a GMM/HMM NSR system, which has then be used with the modified transcription to get new neuro alignments. 

%%%%%%%%%%%%%%%%%%%%%%%%%%%%%%%%%%%%%%%%%%%%%%%%%%%%%%%%%%%%%%%%%%%%%%%%
												%																																			 %
												%					              				Experiments and Results								 %
												%																																			 %
												%%%%%%%%%%%%%%%%%%%%%%%%%%%%%%%%%%%%%%%%%%%%%%%%%%%%%%%%%%%%%%%%%%%%%%%%
\section{Experiments and Results}
\label{sec:Experiments and Results}
												%%%%%%%%%%%%%%%%%%%%%%%%%%%%%%%%%%%%%%%%%%%%%%%%%%%%%%%%%%%%%%%%%%%%%%%%
												%																																			 %
												%					             Experimental setup															 %
												%																																			 %
												%%%%%%%%%%%%%%%%%%%%%%%%%%%%%%%%%%%%%%%%%%%%%%%%%%%%%%%%%%%%%%%%%%%%%%%%
%\subsection{Experimental setup}
%\label{Experimental setup}
%As can be shown in Fig. \ref{fig:ECoG_Array},
\begin{figure}[t!]
%\psfrag{a}[cc][cc][0.7]{Ready}%
%\psfrag{b}[cc][cc][0.7]{Ringo}%
%\psfrag{c}[cc][cc][0.7]{Tiger}%
%\psfrag{d}[cc][cc][0.7]{Go}
%\psfrag{e}[cc][cc][0.7]{To}%
%\psfrag{f}[cc][cc][0.7]{Green}%
%\psfrag{g}[cc][cc][0.7]{Blue}
%\psfrag{h}[cc][cc][0.7]{Red}%
%\psfrag{I}[cc][cc][0.7]{Two}%
%\psfrag{J}[cc][cc][0.7]{Five}%
%\psfrag{K}[cc][cc][0.7]{Seven}%
%\psfrag{L}[cc][cc][0.7]{Now}%
%\psfrag{O}[cc][cc][0.7]{Call sign}%
%\psfrag{M}[cc][cc][0.7]{Color}%
%\psfrag{Q}[cc][cc][0.7]{Number}%
	\centering
		\includegraphics[width=\linewidth]{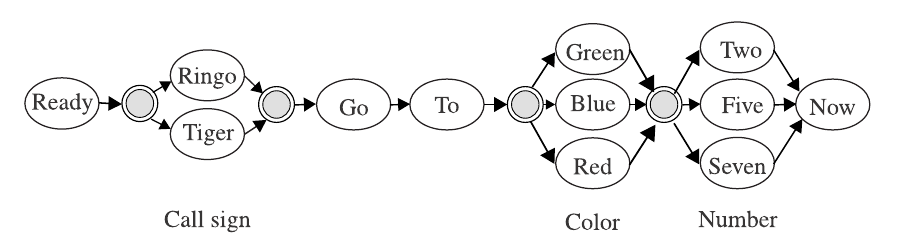}
    \vspace{-5mm}
			\caption{CRM grammar.}
	\label{fig:CRM}
    \vspace{-5mm}
\end{figure}
We have used ECoG signals recorded using a high-density electrode array, which was placed directly over the subject's brain. The subject is an epileptic patient, who had the electrode array superimposed over his brain for clinical purposes prior to a surgery. The array was placed over the auditory cortex to measure the local field potentials of the posterior superior temporal gyrus (pSTG). In \cite{P2009}, it has been shown that this area plays an important role in understanding speech. The high density of electrodes in the electrode array (16$\times$16 electrode array, 4 mm spacing) allows for high resolution recordings in both space and time\footnote{For more details about the ECoG signal recording process, the reader is again referred to \cite{mesgarani2012}}.
%\begin{figure}[t!]
	%\centering
		%\includegraphics[width=0.7\linewidth]{Figures/ECoG_array.eps}
	%\caption{3D reconstruction of a magnetic resonance image (MRI) of the patient's brain, with the electrocorticographic (ECoG) electrode array placed over the auditory cortex. The red and yellow regions are the temporal lobes, which contribute most to speech perception.}
	%\label{fig:ECoG_Array}
%\end{figure}
												%%%%%%%%%%%%%%%%%%%%%%%%%%%%%%%%%%%%%%%%%%%%%%%%%%%%%%%%%%%%%%%%%%%%%%%%
												%																																			 %
												%					              				Datasets					 										 %
												%																																			 %
												%%%%%%%%%%%%%%%%%%%%%%%%%%%%%%%%%%%%%%%%%%%%%%%%%%%%%%%%%%%%%%%%%%%%%%%%
\subsection{Dataset}
\label{Dataset}
The speech corpus used in this study is the Coordinate Response Measure (CRM) corpus \cite{RS2000}. CRM is a small vocabulary corpus (12 words) with a fixed grammar (see Fig. \ref{fig:CRM}). It typically contains phrases like ``Ready Ringo Go To Green Five Now'' spoken by one male and one female speaker. Each speaker utters 12 unique combinations of the two call signs, the three colors, and the three numbers. In addition to the 24 single-speaker signals from the CRM corpus, we have also used 374 utterances from the TIMIT corpus \cite{J1993} to increase the amount of data used for training.

The subject was asked to listen to the single- and mixed-source signals of the CRM utterances in different sessions. Although same utterances were played several times in the different sessions, the recorded ECoG signals were different and stored under different names. Thus, the ECoG corpus contains more ECoG signals than the speech utterances in the CRM corpus. The CRM ECoG corpus contains 155 single-speaker and 61 mixed-speaker signals. The TIMIT ECoG corpus was recorded in one session without replaying. Therefore, there exist the same number of ECoG TIMIT recordings (374 recordings) as of the TIMIT speech utterances. 

Because of the small amount of the CRM data, we have split it into 8 train/test cuts using Monte Carlo cross-validation resampling. For each of the training cuts, we have randomly drawn utterances from the CRM data and augmented them with the 374 TIMIT utterances. The rest of the CRM utterances has been used as a single-source test set. 

We have also used a second test set of mixed-source signals, which were artificially created by adding two signals uttered by the male and female speaker. The two mixed utterances were chosen so that they contain different call signs, colors and numbers.
												%%%%%%%%%%%%%%%%%%%%%%%%%%%%%%%%%%%%%%%%%%%%%%%%%%%%%%%%%%%%%%%%%%%%%%%%
												%																																			 %
												%					              				CPSS    					 										 %
												%																																			 %
												%%%%%%%%%%%%%%%%%%%%%%%%%%%%%%%%%%%%%%%%%%%%%%%%%%%%%%%%%%%%%%%%%%%%%%%%
\subsection{Cocktail Party Simulation Scenario}
\label{CPSS}
The subject first listened to single-speaker utterances from the male and female speaker with different call signs. Then, the subject listened to a mixture of the two utterances. The subject was instructed to indicate the color and number associated with a predefined call sign. The target speaker, who spoke the the intended call sign, was changed randomly in each trial and thus, the subject was required to initially attend to both speakers until the intended call sign was uttered. Then, the subject had to attend to the target speaker to report the color-number combination.
												%%%%%%%%%%%%%%%%%%%%%%%%%%%%%%%%%%%%%%%%%%%%%%%%%%%%%%%%%%%%%%%%%%%%%%%%
												%																																			 %
												%					             Features 													 %
												%																																			 %
												%%%%%%%%%%%%%%%%%%%%%%%%%%%%%%%%%%%%%%%%%%%%%%%%%%%%%%%%%%%%%%%%%%%%%%%%
\subsection{Feature Extraction}
\label{Features}

\subsubsection{Acoustic Features}

We have used 13 mel-frequency cepstral coefficients in addition to their first and second derivative as acoustic feature. A 39-dimensional feature vector has been extracted every 10~ms using a 25~ms Hamming window. The Kaldi speech recognition toolkit \cite{Povey2011} has been used for acoustic feature extraction.

\subsubsection{Neural Features}

Before extracting the neural features, noisy ECoG channels (6 channels) were first identified and rejected \cite{mesgarani2012}. The mean across all good channels at each time point was then subtracted from all remaining 250 healthy channels \cite{mesgarani2012}. Time-frequency analysis was applied to each individual channel by using a filterbank with 42 Gaussian filters with center frequencies ranging from 4 to 250 Hz \cite{edwards2010spatiotemporal}. This results in 42 analytic signals per channel sampled at the raw data sampling frequency 3052~Hz. The speech-related activities are concentrated in the high-gamma frequency bands \cite{Scott2003} ranging from 70 to 160 Hz. Thus, the absolute value of the filterbank output (the analytic amplitude) in this range was averaged. Since the analytic amplitudes are ~Rayleigh distributed \cite{Bendat2000}, the natural logarithm was taken to stabilize the sample variance \cite{Prucnal1987}. Finally, the 250 signals were downsampled to 100 Hz.

In order to extract salient compact neural features, spatial PCA was first applied to the 250-dimensional high-gamma time series to reduce the dimension to 48. The vectors were then rotated using the Varimax rotation approach \cite{kaiser1958varimax} in order to enhance their sparsity properties. Next, convex non-negative matrix factorization (NMF) \cite{ding2010convex} was applied to cluster the correlated neural features while decorrelating the resulting clusters from each other. Finally, the first and second derivatives were cascaded to the main feature vectors to form the final 144-dimensional neural feature vectors.

\begin{table*}[t!]
	\centering{
	\caption{Cross-validation phone error rates of the ASR and NSR system in single- and mixed-speaker scenarios.}
	\label{table:Mixed}
		\begin{tabular}{c|c||cccccccc|c|c}
			\hline
			 \multicolumn{2}{c||}{Cross validation \#}	&    1   &     2   &    3    &    4    &     5   &    6   &   7   &   8  & Average & SD  \\
			\hline
\multirow{2}{*}{ASR} & Single speaker & 4.28 &  3.17 &  2.91 &  2.22 &  3.31 &  6.15 &  4.84 &  5.19 &  4.01 & 1.33  \\
                          & Multi-speaker & 48.86 & 51.21 & 61.29 & 57.47 & 54.08 & 54.65 & 56.61 & 50.00 & 54.27 & 4.18 \\
\hline
\multirow{2}{*}{NSR}     & Single speaker & 57.18 & 56.74 & 55.18 & 57.39 & 56.86 & 56.33 & 58.72 & 55.60 & 56.75 & 1.10 \\
						  & Multi-speaker & 60.03 & 64.55 & 65.59 & 62.23 & 63.10 & 67.50 & 67.54 & 64.54 & 64.54 & 2.58 \\
			\hline
		\end{tabular}
	}
 \vspace{-5mm}
\end{table*}

\subsection{Models}
\label{Models}%left-to-right 
We have used 3-state hidden Markov models (HMMs) to model the acoustic and neural representation of 39 phones. 

The ASR system has been trained as follows: A set of Gaussian mixture model (GMM)/HMM models has been firstly trained, where each HMM emitting state is equipped with 8 Gaussian mixture components. A manually transcribed set of the CRM corpus has been used to initialize the acoustic GMM/HMM models. The model parameters have been conventionally reestimated using the Baum-Welch algorithm. Given the trained GMM/HMM models, the acoustic features, and their corresponding transcription, the forced alignment algorithm has been used to find frame-state alignments, which have been used as labels for training a deep neural network (DNN)/HMM hybrid models. The DNN has 2 hidden layers, each of which consists of 1100 neurons with hyperbolic tangent activation functions. The output layer consists of the 117 (3 states/phone, 39 monophones) monophone states. The input layer of the DNN consists of 250 input units, which have been obtained from splicing 17 input feature vectors followed by linear discriminant analysis (LDA) \cite{kolossa2013noise} for dimension reduction. Stochastic gradient descent has been used to estimate the DNN parameters. The network has been trained for a total of 20 epochs. 

We have used a biphone language model, which has been trained using the transcription of the training set. 

The NSR system has been trained similarly to the ASR system described above. However, the GMM/HMM NSR system has been trained using the frame-state alignments obtained from the DNN/HMM ASR system. No realignment steps have been conducted after each mixture splitting and maximization step in the Baum-Welch algorithm. The GMM/HMM NSR system has been used to obtain new alignments for training a DNN/HMM NSR system. As described in Section \ref{sec:ANA}, additional silences have been embedded in the GMM/HMM NSR training lattices to account for the transcription mismatch between the acoustic and neural signals. A silence skip probability of 0.9 has been empirically chosen. 

												%%%%%%%%%%%%%%%%%%%%%%%%%%%%%%%%%%%%%%%%%%%%%%%%%%%%%%%%%%%%%%%%%%%%%%%%
												%																																			 %
												%					             					Results																 %
												%																																			 %
												%%%%%%%%%%%%%%%%%%%%%%%%%%%%%%%%%%%%%%%%%%%%%%%%%%%%%%%%%%%%%%%%%%%%%%%%
\subsection{Results}
\label{Results}
Table \ref{table:Mixed} shows the phone error rate (PER) of the 8 Monte Carlo cross validation experiments, their average, and their standard deviation. As can be seen, initializing the ASR models using the manually transcribed set of data has significantly reduced the average PER of the ASR and consequently the NSR system compared to the results in \cite{Chang2015}. However, the results show similar behavior to that observed in \cite{Chang2015}: The ASR performs quite well in the single-speaker scenario. However, this performance degrades dramatically in the mixed-source scenario with a 93\% relative increase in the average PER. For the NSR system, the average performance degradation from the single-source to the mixed-source scenario is only 12\%, which conforms to the fact that humans tend to be more robust to interfering signals than ASR systems.    

Table \ref{table:CHMM} shows the PER of the NSR system when two different frame-state alignments were used. The first set of frame-state alignments has been obtained from the DNN/HMM ASR system, while the second one has been estimated using the GMM/HMM NSR system with the additional embedded silences. As can be seen, using the GMM/HMM-NSR-based alignments gives 4.5\% relative PER reduction compared to using the DNN/HMM-ASR-based alignments. 

\begin{table}[t!]
	\centering{
	\caption{Phone error rate (PER) of the DNN/HMM NSR system trained with different frame-state alignments.}
	\label{table:CHMM}
		\begin{tabular}{c|c}
			\hline
			Frame-state alignments    &  PER  \\
			\hline
			 DNN/HMM ASR alignments & 59.42 \\  
			 GMM/HMM NSR alignments & 56.75 \\
			\hline
		\end{tabular}
	}
\vspace{-5mm}
\end{table}
 
%This is because of the additional silences that have been automatically added while aligning the data using the GMM/HMM NSR system. As mentioned in Section \ref{sec:ANA}, these silences were added in the training lattices with a skip probability of 0.9. As can be seen, this has relatively reduced the  PER by about 4.5\%.
%%%%%%%%%%%%%%%%%%%%%%%%%%%%%%%%%%%%%%%%%%%%%%%%%%%%%%%%%%%%%%%%%%%%%%%%
												%																																			 %
												%					              			Conclusions															 %
												%																																			 %
												%%%%%%%%%%%%%%%%%%%%%%%%%%%%%%%%%%%%%%%%%%%%%%%%%%%%%%%%%%%%%%%%%%%%%%%%
\section{Conclusions\vspace{-1mm}}
\label{sec:Conclusions}
In this paper, we have compared the performance of an ASR and NSR system in a single- and mixed-source scenario. Although the ASR system performance is better than that of the NSR system in the single-source case, the relative phone error rate increment of the ASR system in the mixed-source scenario is far worse than that of the NSR system. This shows the utility of NSR systems to \emph{objectively} prove the robustness of humans in recognizing speech in multitalker environments compared to ASR systems. We have significantly improved our previous ASR and NSR results in \cite{Chang2015} by initializing the ASR system using manually transcribed data. Further, we have improved the NSR results by considering the transcription mismatch between the acoustic and ECoG signals.

A natural extension of this work is to further investigate the alignment mismatch between the acoustic and ECoG signals, which may allow us to answer questions like: how do acoustic environments influence the amount of speech signal processing applied by human brains? This might be another step forward to our ultimate goal of building an ASR system that possesses the capability of human listeners for speech recognition under noisy acoustic conditions.

%\nocite{Abdelaziz2012}
%\newpage
\bibliographystyle{IEEEtran}
\bibliography{litlist}

\end{document}